# Constructing a bifunctional platform based on Mn$^{2+}$-doped Mg$_2$Y$_8$(SiO$_4$)$_6$O$_2$ phosphors for multi-parameter optical thermometry and manometry

*Zhiyu Pei, Shuailing Ma\*, Maja Szymczak, Lukasz Marciniak\*, Tian Cui, Laihui Luo, Peng Du\**

Z. Pei, S, Ma, T. Cui, L. Luo, P. Du

School of Physical Science and Technology, Ningbo University, 315211 Ningbo, Zhejiang, China

\*E-mail: mashuailing@nbu.edu.cn (S. Ma); dupeng@nbu.edu.cn or dp2007good@sina.com (P. Du)

M. Szymczak, L. Marciniak

Institute of Low Temperature and Structure Research, Polish Academy of Sciences, Okólna 2, 50-422 Wrocław, Poland

\*E-mail: l.marciniak@intibs.pl (L.Marciniak)

**ABSTRACT**

**Series of the Mn$^{2+}$-doped Mg$_2$Y$_8$(SiO$_4$)$_6$O$_2$ phosphors were synthesized. Upon excitation at 408 nm, these phosphors exhibited intense orange emission originating from Mn$^{2+}$, with concentration quenching observed beyond *x* = 0.07, and they also demonstrated excellent thermal stability. For optical thermometry, two independent parameters, emission band centroid (λ) and lifetime, were employed as thermal indicators, yielding sensitivities of d$\lambda$/d*T* = 0.053 nm K$^{-1}$ and *S$_R$* = 0.86% K$^{-1}$, respectively. High-pressure *in-situ* X-ray diffraction revealed that the phosphors retained structural integrity under compression, accompanied by a progressive lattice contraction. With increasing pressure (0.13-10.89 GPa), a spectral red-shift was observed, corresponding to a pressure sensitivity of d$\lambda$/d*p* = 4.75 nm GPa$^{-1}$. Additionally, pressure-dependent shifts in color coordinates allowed the development of a colorimetric manometric response, achieving a relative sensitivity of**

**3.27% GPa$^{-1}$. Remarkably, the pressure-induced spectral shift of Mn$^{2+}$ emission, characterized by low thermal cross-sensitivity, enabled a highly reliable ratiometric manometric strategy, with a relative sensitivity of 72% GPa$^{-1}$. Notably, the system delivered the highest TIMF reported to date above 3 GPa, peaking at 1940 K GPa$^{-1}$ at 7 GPa. These results position Mn$^{2+}$-doped Mg$_2$Y$_8$(SiO$_4$)$_6$O$_2$ phosphors as a highly promising bifunctional material for next-generation, multi-parameter optical sensing applications under extreme conditions.**

## 1. Introduction

Temperature and pressure, as two fundamental physical parameters, exhibit great influence on the industrial manufacture and instinct characteristics, *i.e.*, electronic structure, bond length, phase composition, luminescence, *etc.*, of materials, and thereby, their accurate monitoring is very important.[1,2] Compared with the conventional measuring routes based on the physical contact-based manometers and thermometers, the remote optical monitoring methods for calibrating temperature and pressure have drawn considerable interests in the view of its advantages of fast response, non-invasive detection, high spatial resolution, and the feasibilities in some harsh conditions, *i.e.*, oil/gas industry, power station, micro or nano-sized objects, *etc.*[3-5] As is known, through analyzing the responses of the spectroscopy characteristics of the luminescent materials to the extreme conditions (*i.e.*, pressure and temperature), the remote optical sensing is able to be realized.[6,7] In terms of the luminescence thermometry, most of the prior reports took the advantage of the luminescence intensity ratio of two emission bands from rare-earth or transition metal ions as the temperature indicator.[8,9] Although this technique can result in high sensitive optical thermometers and exhibits high resistance to external conditions (*i.e.*, pump power and dopant content), it suffers from the complex interactions between medium and light, which will lead to the errors in the temperature readouts.[10] Thereby, to settle this deficiency, an effective strategy, *i.e.*, using diverse spectroscopic parameters (*i.e.*, lifetime, band width, emission band centroid, *etc*.) as the metrological parameters, is required to be proposed so as to design new optical thermometers with high accuracy and sensitivity.

Currently, to meet the requirements of the scientific research, the diamond anvil cell (DAC) has been extensively adopted to create the hydrostatic pressure, in which the ruby is commonly used to remotely detect the generated high-pressure. However, the pressure sensitivity (*i.e.*, d$\lambda$/d$p$ = 0.365 nm GPa$^{-1}$) of ruby is not high enough to satisfy the demands of technological and scientific development. In an attempt to solve this problem, researchers have developed series of optical manometers based on the luminescent materials, such as Li$_4$SrCa(SiO$_4$)$_2$:Eu$^{2+}$, Sr$_2$[Mg$_{0.9}$Li$_{0.1}$Al$_{4.9}$Si$_{0.1}$N$_7$]:Eu$^{2+}$, Ba$_3$Lu(BO$_3$)$_3$:Ce$^{3+}$, NaMgBO$_3$:Ce$^{3+}$, Zn$_3$Ga$_2$GeO$_8$:Mn$^{4+}$, La$_3$Mg$_2$SbO$_9$:Mn$^{4+}$, *etc*.[11-16] Compared with that of the ruby, the manometric properties of these luminescent materials had been improved, suggesting that the development of the new luminescent material is an efficient method to design highly sensitive optical manometers. However, when pressure imaging is required, analysis based on the spectral position of a single emission band becomes impractical due to its time-consuming nature. In such cases, the recently introduced ratiometric approach can be employed.[17] To date, this method has been reported only for Cr$^{3+}$ and Ce$^{3+}$.[18-20] As is known, to develop highly pressure response optical manometer, selecting proper activators, specifically, whose luminescence features are sensitive to the crystallographic environment, is required. Nowadays, considerable interests in transition metal ions, such as Bi$^{3+}$, Mn$^{2+}$, Fe$^{3+}$, *etc*., have been gained on account of their abundant emissions, *i.e.*, from ultraviolet to near-infrared.[21-23] Among them, Mn$^{2+}$, which exhibits the *3d$^5$* electronic configuration, is frequently adopted as activator due to its featured emission arises from the $^4T_1(^4G) \rightarrow {^6}A_1$ transition within the *3d* orbit.[24,25] As has been well documented, the *3d* orbital transition of Mn$^{2+}$ is significantly influenced by the crystal field, that is, when Mn$^{2+}$ takes up site with weak crystal field, it will emit green emission, otherwise, a red-orange emission will be produced when it occupies the site with strong crystal field.[26,27] Thus, the luminescence of Mn$^{2+}$ can be manipulated through adjusting its surrounding environment. Interestingly, when the high-pressure is applied to the object, the lattice will be compressed, which shortens the distance among the atoms, contributing to the modified crystal field.[28,29] As a result, via utilizing the high-pressure engineering, the spectral characterizations of $Mn^{2+}$ are able to be regulated, enabling its applications in remote optical manometry. Thereby, the investigation on the responses of the $Mn^{2+}$ luminescence to pressure so as to develop high-sensitive optical manometers will be interesting.

Considering the practical applications, the designed luminescent materials are required to present good luminescence properties. Since the $Mn^{2+}$ luminescence is determined by the coordinate environment, choosing a proper host is the simplest pathway to design the highly-efficient $Mn^{2+}$-doped phosphors. At present, silicate oxyapatites with the chemical formula of $M_{10}(SiO_4)_6O_2$ (M = Na, Li, Ba, Ca, Sr, Y, Gd, La, *etc*.) have been extensively studied as the luminescent hosts for rare-earth and transition metal ions as a result of their remarkable advantages including proper phonon energy, high physicochemical stability, unique structure and high rigidity.[30,31] Notably, via the utilization of $Mg_2Y_8(SiO_4)_6O_2$ compound as host, several kinds of phosphors, *i.e.*, $Mg_2Y_8(SiO_4)_6O_2:Ce^{3+}$, $Mg_2Y_8(SiO_4)_6O_2:Eu^{3+}$, $Mg_2Y_8(SiO_4)_6O_2:Tb^{3+}$, *etc*., had been developed for solid-state lighting application.[32-34] However, as far as we know, the investigation on the spectral regulation of the $Mn^{2+}$-doped $Mg_2Y_8(SiO_4)_6O_2$ phosphors induced by high-pressure has never been performed yet. Here, series of $Mg_{2-2x}Y_8(SiO_4)_6O_2:2xMn^{2+}$ (*i.e.*, $Mg_2Y_8(SiO_4)_6O_2:2xMn^{2+}$) phosphors were prepared, of which their phase structure, electronic structure, morphology, elemental compositions and luminescence were analyzed detailly. High thermal and manometric variability of spectroscopic parameters of $Mn^{2+}$-doped $Mg_2Y_8(SiO_4)_6O_2$ enables the development of multimodal bifunctional temperature-pressure sensor. Moreover, as demonstrated in this study, the high sensitivity of the $Mn^{2+}$ emission band position in $Mn^{2+}$-doped $Mg_2Y_8(SiO_4)_6O_2$ to pressure changes, combined with its negligible thermal dependence, enabled the development of a ratiometric luminescent manometer with a record-high thermal insensitivity in the high-pressure regime above 3 GPa. The results demonstrated the $Mn^{2+}$-doped $Mg_2Y_8(SiO_4)_6O_2$ phosphors with good luminescence properties are promising bifunctional platforms for multi-parameter optical manometry and thermometry.

## 2. Results and Discussion

The evolution of the phase structure of the resulting phosphors at various $Mn^{2+}$ contents was characterized via XRD technique. As displayed in Figure 1(a), all the samples exhibit the similar diffraction profiles and they are in accordance well with the reference pattern for hexagonal phase of $Mg_2Y_8(SiO_4)_6O_2$ (ICSD#195494), in which any additional diffraction peaks from secondary phase are not observed, implying that the doping of $Mn^{2+}$ does not alter the phase of the studied samples in the analyzed dopant concentration range. To comprehend the impact of $Mn^{2+}$ doping on the crystal structure of the final products in depth, the typical Rietveld XRD refinements of the $Mg_2Y_8(SiO_4)_6O_2:0.02Mn^{2+}$ and $Mg_2Y_8(SiO_4)_6O_2:0.14Mn^{2+}$ phosphors were executed, as presented in Figure 1(b) and 1(c), respectively. Here, these calculated diffraction profiles coincide well with the experimental data, suggesting that synthesized phosphors possess pure hexagonal phase. Although, the phase transition does not take place, it is found that the lattice parameters (*i.e.*, $a = b$, $c$ and volume) are changed as $Mn^{2+}$ adds, *i.e.*, their values are increased with the increment of doping content (see Table S1), which is assigned to the larger-sized $Mn^{2+}$ replaces the smaller-sized $Mg^{2+}$ in $Mg_2Y_8(SiO_4)_6O_2$ host lattices. The crystal structure of the $Mg_2Y_8(SiO_4)_6O_2$ compound (*i.e.*, P$6_3$/m (176) space group) is shown in Figure 1(d). As disclosed, both Mg and Y atoms occupy two crystallographic sites, namely, (i) 4f site with $C_3$ symmetry surrounded by nine oxygen atoms, resulting in the formation of $[MgO_9]$ and $[YO_9]$ polyhedrons, and (ii) 6h site with $C_s$ symmetry coordinated by seven oxygen atoms, leading to the generation of the $[MgO_7]$ and $[YO_7]$ polyhedrons. Generally, to form a new solid solution via doping engineering, the ionic radius difference ($D_r$) between dopant and replaced ions should be small enough (*i.e.*, $D_r < 30\%$) and it can be determined according to the well know formula: $D_r = |(R_1 - R_2)/R_2| \times 100\%$ (where the ionic radii of replaced ions and dopant are labelled by $R_1$ and $R_2$, respectively).[35] As a result, the $D_r$ values of the $Mg^{2+}/Mn^{2+}$ combination are estimated to be 11.2% and 8.06%, respectively, when the coordinate number is 7 and 9, which are all less than 30%, revealing that the designed phosphors can be formed, coinciding well with the XRD results.

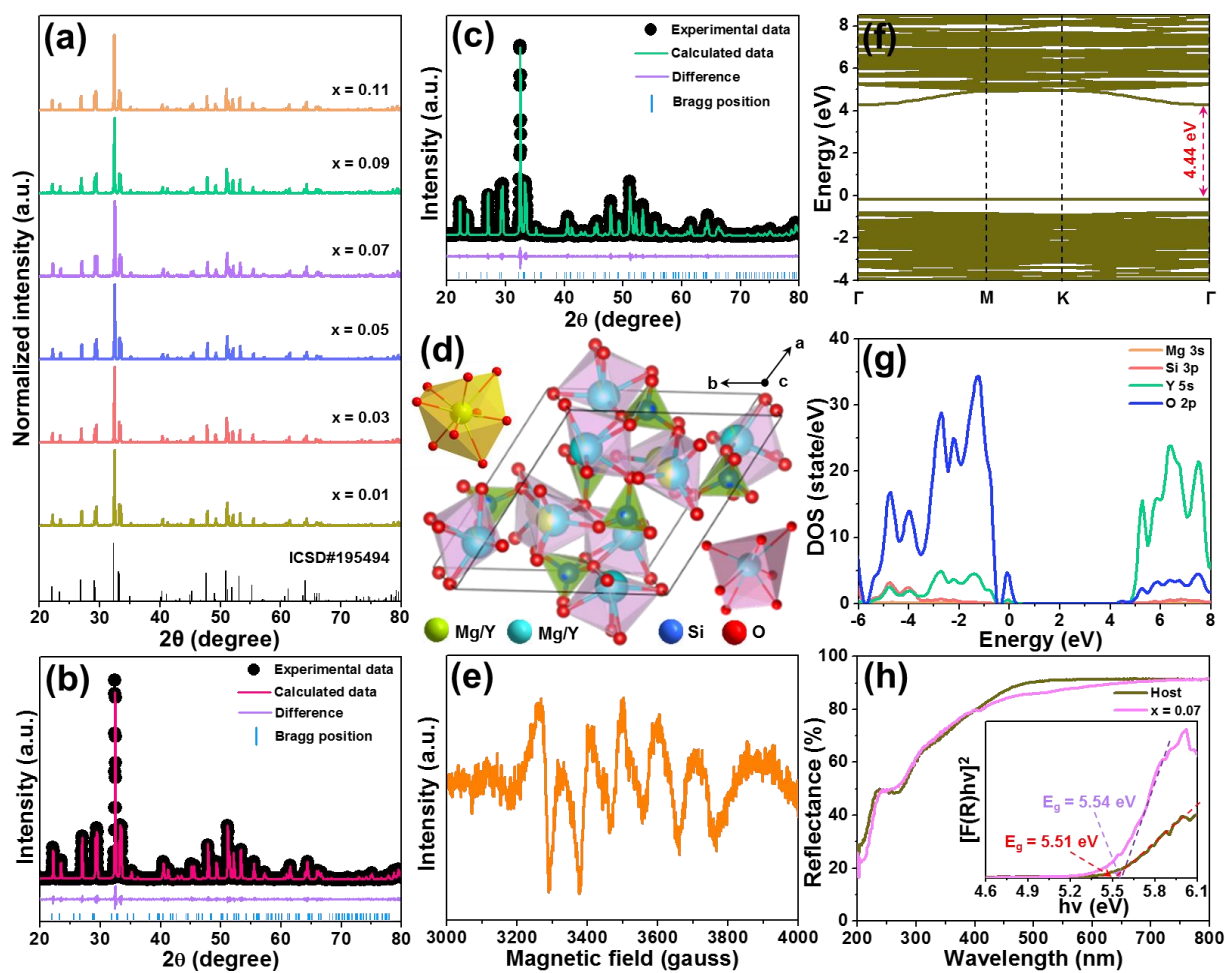

**Figure 1** (a) XRD patterns of the $Mg_2Y_8(SiO_4)_6O_2:2x Mn^{2+}$ phosphors with different dopant concentration. Rietveld XRD refinements of the (b) $Mg_2Y_8(SiO_4)_6O_2:0.02Mn^{2+}$ and (c) $Mg_2Y_8(SiO_4)_6O_2:0.14Mn^{2+}$ phosphors. (d) Crystal structure of the $Mg_2Y_8(SiO_4)_6O_2$ compound. (e) EPR spectrum of the $Mg_2Y_8(SiO_4)_6O_2:0.14Mn^{2+}$ phosphors. (f) Electronic structure and (g) DOS of the $Mg_2Y_8(SiO_4)_6O_2$ compound. (h) Diffuse reflectance spectra of the $Mg_2Y_8(SiO_4)_6O_2$ and $Mg_2Y_8(SiO_4)_6O_2:0.14Mn^{2+}$ phosphors. Inset shows the calculated $E_g$ values.

For the purpose of analyzing the exact elemental compositions as well as their valences in the designed compound, the X-ray photoelectron spectroscopy (XPS) test was performed (Figure S1). The full XPS survey spectrum, shown in Figure S1(a), manifests that the resulting phosphors consist of the $Mg^{2+}$, $Y^{3+}$, $Si^{4+}$, $O^{2-}$ and $Mn^{2+}$ elements. The high-resolution XPS spectrum of the $Mg^{2+}$ 1s is illustrated in Figure S1(b) and it contains an intense peak at 1303.7 eV. The high-resolution XPS spectrum of $Y^{3+}$ 3d (Figure S1(c)) consists of two bands at 157.6 and 159.7 eV corresponding to $Y^{3+}$ $3d_{3/2}$ and $Y^{3+}$ $3d_{1/2}$, respectively.[36] Figure S1(d) describes the high-resolution XPS spectrum of $Si^{4+}$ 2p and a strong band at 101.3 eV, which is assigned to the $Si^{4+}$ $2p_{3/2}$, is observed.[27] Moreover, a single band with the binding energy of 530.4 eV is detected in the high-resolution XPS spectrum of $O^{2-}$ 1s (see Figure S1(e)). Furthermore, the high-resolution XPS spectrum of $Mn^{2+}$ 2p is composed of two peaks at 641.5 and 653.2 eV, which are attributed to $Mn^{2+}$ $2p_{3/2}$ and $Mn^{2+}$ $2p_{1/2}$, respectively, as displayed in Figure S1(f).[27] To further verify the existence of $Mn^{2+}$ in studied samples, the electron paramagnetic resonance (EPR) spectrum of the $Mg_2Y_8(SiO_4)_6O_2$:0.14$Mn^{2+}$ phosphors was recorded and demonstrated in Figure 1(e). As it is well known, the high-spin $Mn^{2+}$ possesses five unpaired electrons, which makes its EPR spectrum contains sextet lines owing to the interaction between electron spin and its nuclear spin (*i.e.*, I = 5/2).[37] Evidently, the measured EPR spectrum exhibits six significant peaks (see Figure 1(e)), implying that the $Mn^{2+}$ is formed in the designed compounds, which is also homogenously distributed in the lattices without forming $Mn^{2+}$ dimers.[37] These results indicate that the $Mn^{2+}$-doped $Mg_2Y_8(SiO_4)_6O_2$ phosphors were successfully synthesized.

To clarify the electronic structure of $Mg_2Y_8(SiO_4)_6O_2$ compound, a theoretical calculation was carried out via using the CASTEP program. As demonstrated in Figure 1(f), the $Mg_2Y_8(SiO_4)_6O_2$ compound shows a direct band gap of $E_g$ = 4.44 eV, in which the bottom of conduction band and the top of the valence band all situate at Γ point. Moreover, it is shown in the density of state (DOS) that both conduction band minimum and valence band maximum are all mainly contributed by the Y 5s and O 2p orbitals (see Figure 1(g)). Figure 1(h) illustrates the diffuse reflectance spectra of the $Mg_2Y_8(SiO_4)_6O_2$ and $Mg_2Y_8(SiO_4)_6O_2:0.14Mn^{2+}$ phosphors. In comparison, the $Mg_2Y_8(SiO_4)_6O_2:0.14Mn^{2+}$ phosphors presents an extra band centered at 408 nm associated with the $^6A_1 \rightarrow {}^4A_1(^4G), {}^4E(^4G)$ electronic transition of $Mn^{2+}$, as presented in Figure 1(h). Furthermore, through utilizing the following formula to analyze the diffuse reflectance spectrum, the $E_g$ value of the studied samples can be estimated:[6,27]

$$F(R) = (1-R)^2/2R \tag{1}$$

$$h\nu F(R) = A(h\nu - E_g)^n \tag{2}$$

where $h\nu$ represents the photon energy, $A$ is constant, $R$ corresponds to the reflectivity and $n$ value is decided by the semiconductor category. Herein, $n$ equals to 1/2 since $Mg_2Y_8(SiO_4)_6O_2$ is attributed to the direct band gap semiconductor. Thus, the plots of $[F(R)h\nu]^2$ vs. $h\nu$ are obtained and shown in the inset of Figure 1(h). Clearly, the $E_g$ value of the $Mg_2Y_8(SiO_4)_6O_2$ and $Mg_2Y_8(SiO_4)_6O_2:0.14Mn^{2+}$ phosphors are 5.51 and 5.54 eV, respectively, which suggests that the introduction of $Mn^{2+}$ has little impact on the $E_g$ value. Notably, the calculated and experimental $E_g$ values exhibit an obvious difference, which may be associated with the fact that the density functional theory approximation always underestimates the $E_g$ value.

To analyze the morphology the synthesized samples, the SEM and TEM characterizations were adopted and the corresponding results are shown in Figure 2. From the SEM images (see Figure 2(a)-2(c)), one knows that the final products are constituted of irregular microparticles, of which the particle size and shape are independent on dopant content. Moreover, the TEM image shown in Figure 2(d) also reveals the particles within the studied samples exhibit a micro-scale size. Figure 2(e) displays the high-resolution TEM image of the $Mg_2Y_8(SiO_4)_6O_2:0.14Mn^{2+}$ phosphors, in which distinct lattice fringes are detected and their adjacent distance is found to be 3.57 Å pertaining to the (021) crystal plane of the $Mg_2Y_8(SiO_4)_6O_2$ host. Furthermore, the selected-area electron diffraction (SAED) pattern comprises many bright dots (Figure 2(f)), manifesting the single crystalline nature of the particles. Additionally, the energy dispersive spectroscopy (EDS) analysis also indicates that the Mg, Y, Si, O and Mn elements exist in the designed phosphors (Figure 2(g)), and they distribute uniformly throughout the whole microparticles, as verified by the elemental mapping results (see Figure 2(h)-2(m)).

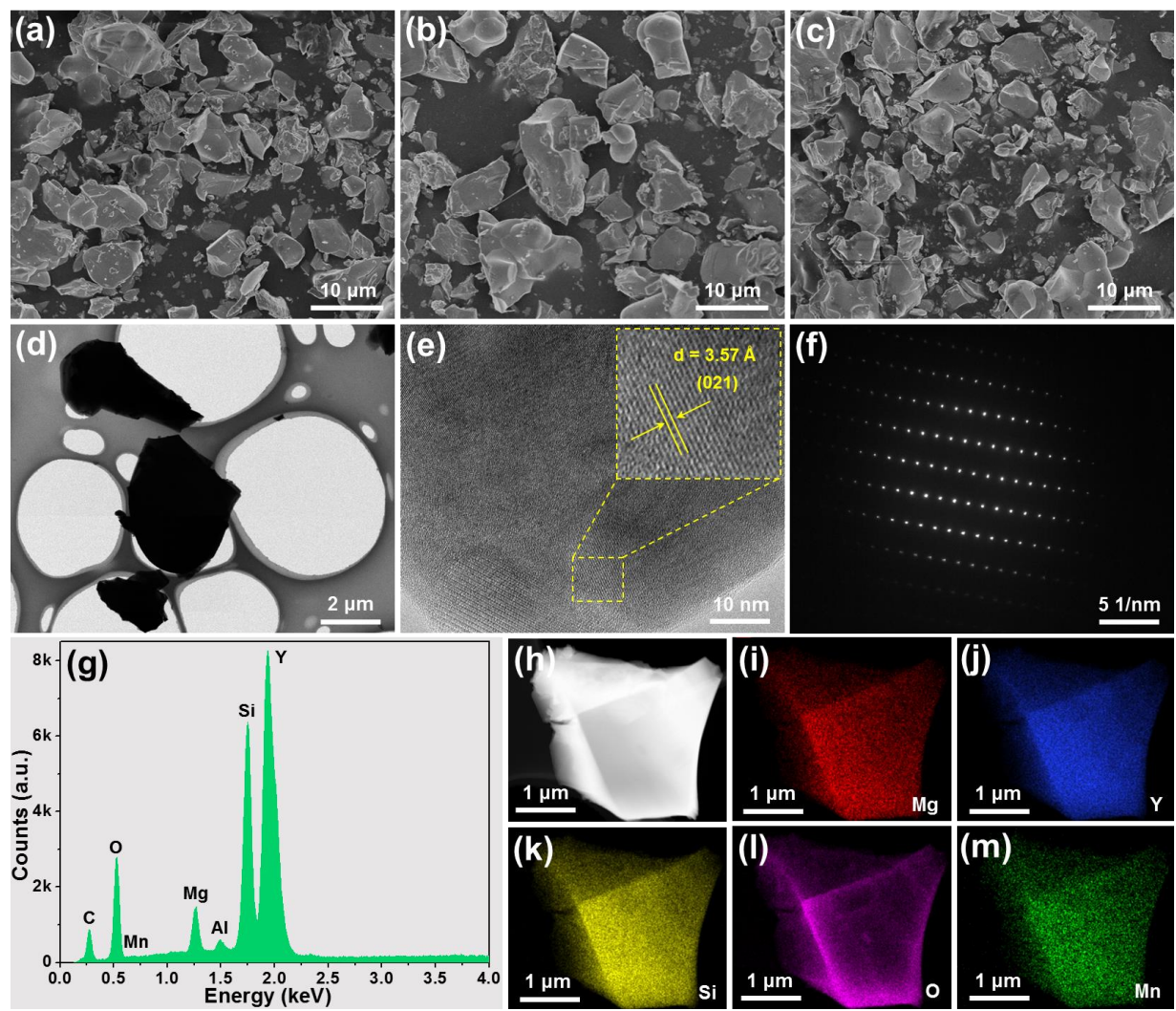

**Figure 2** SEM images of the $Mg_2Y_8(SiO_4)_6O_2$:$2x$Mn$^{2+}$ phosphors with the dopant content of (a) $x$ = 0.01, (b) $x$ = 0.07 and (c) $x$ = 0.11. (d) TEM image, (e) High-resolution TEM image, (f) SAED pattern, (g) EDS spectrum and (h)-(m) Elemental mapping of the $Mg_2Y_8(SiO_4)_6O_2$:0.14Mn$^{2+}$ phosphors.

For the aim of revealing the luminescence features of the studied samples, their excitation and emission spectra were tested. Monitored at 609 nm, it can be seen that the excitation spectrum of the $Mg_2Y_8(SiO_4)_6O_2$:0.14Mn$^{2+}$ phosphors comprises four bands at approximately 345, 366, 408 and 468 nm, which originate from the $d$-$d$ transitions of Mn$^{2+}$ from $^6A_1$ ground state to $^4E(^4D)$, $^4T_2(^4D)$, $^4A_1(^4G)$,$^4E(^4G)$ and $^4T_2(^4G)$ levels, respectively, as shown in Figure 3(a).[24-26] In view of the strongest intensity of the excitation band at 408 nm, it was selected as the excitation wavelength to monitor the emission spectra of the studied samples. The emission spectra of the $Mg_2Y_8(SiO_4)_6O_2$:0.14$Mn^{2+}$ phosphors contain an asymmetrical broad band centered at 608 nm (*i.e.*, $^4T_1(^4G) \rightarrow {}^6A_1$ transition), which is assigned to the two nonequivalent crystallographic sites occupied by the $Mn^{2+}$ in the host lattices (*i.e.*, 4f and 6h sites). The deconvolution of the emission spectrum reveals the presence of two bands centered at at 603 and 648 nm (see Figure 3(a)). As disclosed in Figure 3(b), the emission intensity of the analyzed phosphor increases gradually with the increment of $Mn^{2+}$ content and the optimum value is found at $x = 0.07$, whereas the concentration quenching effect appears with further rising the doping content. To understand the corresponding concentration quenching mechanism, it is necessary to analyze the critical distance ($R_c$) between the dopants. Here, the $R_c$ value of $Mn^{2+}$ in $Mg_2Y_8(SiO_4)_6O_2$ host lattice is estimated to be 18.94 Å (for the detailed calculations, see Supporting Information), which is larger than 5 Å, suggesting that the electric multipolar interaction contributes to the concentration quenching mechanism. Furthermore, via investigating the relation between fluorescence intensity ($I$) and dopant content ($x$), more details of the concentration quenching mechanism can be achieved, as described below:[38]

$$\log\left(\frac{I}{x}\right) = A - \frac{\theta}{3}\log(x) \qquad (3)$$

where $A$ refers to the constant and the values of $\theta$ equal to 6, 8 and 10 correspond to electric dipole-dipole, dipole-quadrupole and quadrupole-quadrupole interactions, respectively. Figure S2(a) shows the plot of $\log(I/x)$ *vs.* $\log(x)$ for the $Mg_2Y_8(SiO_4)_6O_2$:$2x$$Mn^{2+}$ phosphors. Through linearly fitting the experimental data, one knows that the slope of the fitted line is $-\theta/3 = -3.28$, implying that the $\theta$ value is 9.84. Since the calculated $\theta$ value approaches to 10, the involved concentration quenching mechanism is triggered by the electric quadrupole-quadrupole interaction.



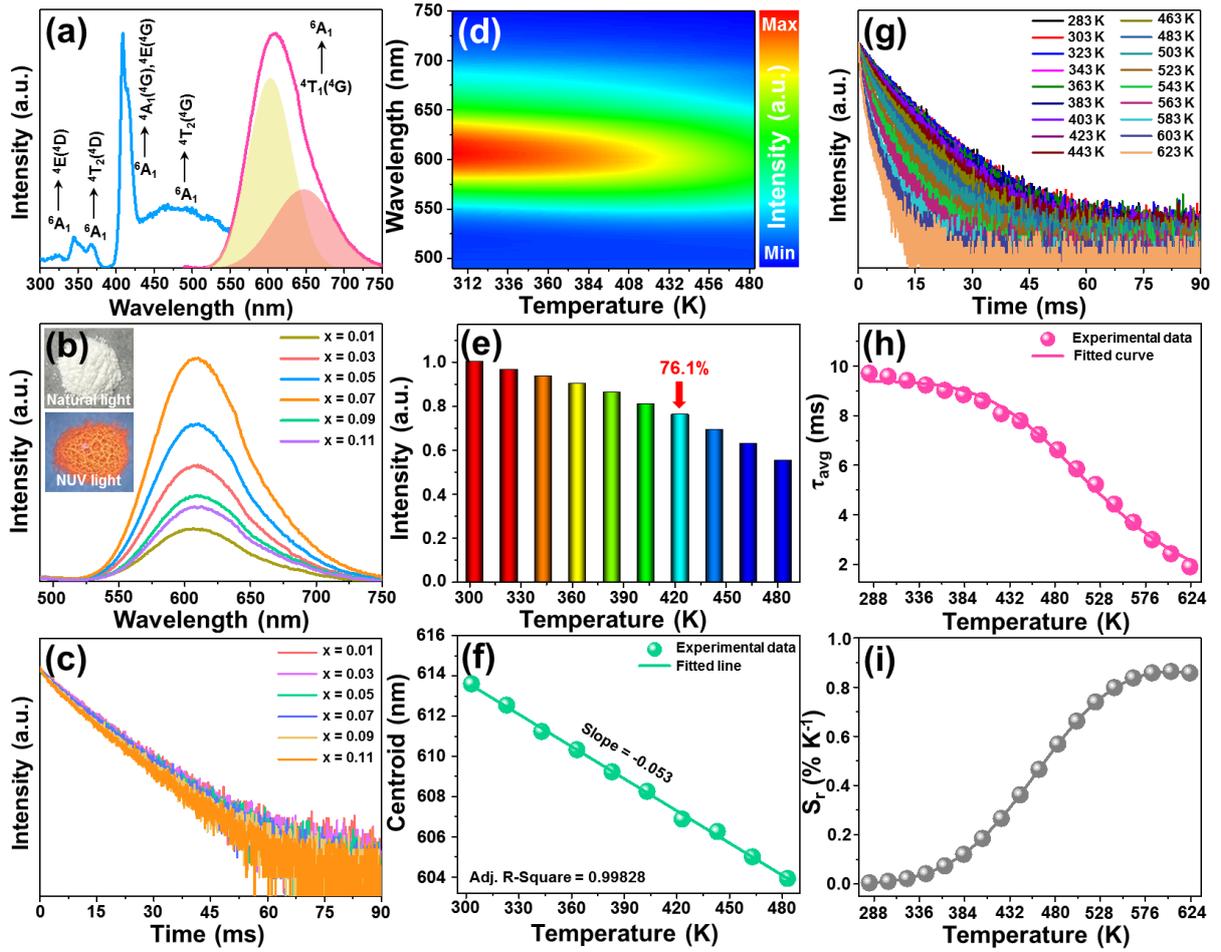

**Figure 3** (a) Excitation and emission spectra of the $Mg_2Y_8(SiO_4)_6O_2$:0.14$Mn^{2+}$ phosphors. (b) Emission spectra and (c) decay curves of the $Mg_2Y_8(SiO_4)_6O_2$:2$x$$Mn^{2+}$ phosphors. Temperature-dependent (d) emission spectra, (e) Normalized fluorescence intensity, (f) emission band centroid, (g) decay curves, (h) lifetime and (i) $S_r$ value of the $Mg_2Y_8(SiO_4)_6O_2$:0.14$Mn^{2+}$ phosphors.

Irradiated by the near-ultraviolet light, it is found that the resulting phosphors can emit visible orange light and its color coordinate is (0.565,0.433), as presented in Figure 3(b) and S2(b). Figure 3(c) shows the room temperature decay curves of the $Mg_2Y_8(SiO_4)_6O_2$:2$x$$Mn^{2+}$ phosphors ($\lambda_{exc}$ = 408 nm, $\lambda_{em}$ = 608 nm). Evidently, the recorded decay curves exhibit a second-order exponential decay, further confirming that the $Mn^{2+}$ occupies two different crystallographic sites in the $Mg_2Y_8(SiO_4)_6O_2$ host lattices, and they can be fitted by the following formula:

$$I(t) = A_1\exp(-t/\tau_1) + A_2\exp(-t/\tau_2) \qquad (4)$$



where the fluorescence intensity at time $t$ is labeled by I($t$), $A_i$ ($i$ = 1, 2) denotes constant and $\tau_i$ ($i$ = 1, 2) describes the lifetime. Moreover, the average lifetime ($\tau_{avg}$) can be evaluated by utilizing the following function:

$$\tau_{avg} = (A_1\tau_1^2 + A_2\tau_2^2)/(A_1\tau_1 + A_2\tau_2) \tag{5}$$

Accordingly, the $\tau_{avg}$ of $Mn^{2+}$ in the designed phosphors is calculated to be 9.85, 9.83, 9.66, 9.64, 9.23 and 8.87 ms, respectively, when the $x$ = 0.01, 0.03, 0.05, 0.07, 0.09 and 0.11. Apparently, the $\tau_{avg}$ decreases as doping concentration arises, manifesting the existence of concentration quenching in the synthesized phosphors.

To get deeper insight into the luminescence behaviors of the developed compounds, the temperature-dependent emission spectra of the $Mg_2Y_8(SiO_4)_6O_2$:0.14$Mn^{2+}$ phosphors were examined (Figure 3(d)). As temperature rises, it is found that the emission intensity decreases gradually, which is assigned to the thermal-induced nonradiative depopulation of the excited state of $Mn^{2+}$ ions. Notably, when the temperature reaches up to 423 K, the fluorescence intensity still remains 76.1% of its initial value at 303 K (see Figure 3(e)), demonstrating that the designed phosphors possess good thermal stability, which enables its feasibility in high-temperature conditions. For analyzing the thermal quenching effect in depth, the involved activation energy ($\Delta E$) is estimated utilizing the following function:[39,41]

$$I(T) = \frac{I_0}{1+A\exp(-\Delta E/kT)} \tag{6}$$

where the integral emission intensities at initial and recorded temperature are labelled by $I_0$ and $I$(T), respectively, $A$ is constant and $k$ stands for the Boltzmann coefficient. As demonstrated in Figure S3, the $\Delta E$ value of $Mn^{2+}$ in the $Mg_2Y_8(SiO_4)_6O_2$ host lattices is 0.255 eV. Interestingly, aside from the quenched luminescence at high-temperature, a spectral blue shift is also observed in the synthesized phosphors (Figure 3(d)), which results from the expansion of the crystal lattice, leading to the shrinkage of the band gap.[42] Specifically, with increasing the temperature from 303 to 483 K, it is found that the emission band centroid (*i.e.*, $\lambda$) changes from 613.6 to



603.9 nm, as illustrated in Figure 3(f), of which the relation between emission band centroid and temperature is linear, resulting in the temperature sensitivity of $d\lambda/dT = 0.053$ nm K$^{-1}$. These results demonstrate that the spectral behaviors of the prepared phosphors are able to be regulated via high-temperature engineering, endowing their feasibilities in optical temperature monitoring.

Figure 3(g) presents the temperature-associated decay curves of the Mg$_2$Y$_8$(SiO$_4$)$_6$O$_2$:0.14Mn$^{2+}$ phosphors, in which the monitoring and excitation wavelengths are 609 and 408 nm, respectively. According to the recorded decay curves, one knows that the lifetime of the Mg$_2$Y$_8$(SiO$_4$)$_6$O$_2$:0.14Mn$^{2+}$ phosphors deceases from 9.71 to 1.91 ms, which results from the thermal quenching effect, as temperature elevates from 283 to 623 K. As described in previous reports,[43,44] it is evident that the following function can be used to describe the temperature-dependent lifetime:

$$\tau = \frac{\tau_0}{1 + A\exp(-\Delta E/kT)} \quad (7)$$

where the lifetime of the studied samples at $T = 0$ K is labeled by $\tau_0$ and $\tau$ refers to the average decay time at studied temperature, $A$ is constant, $\Delta E$ and $k$ exhibit the same meanings as described in Eq. (6). Through analyzing the experimental data (see Figure 3(h)), one achieves that relation between lifetime and temperature in the studied samples is $\tau = 9.39/[1 + 3492.33\exp(-4336.28/T)]$. In order to investigate the thermometric properties of the resultant phosphors, its $S_R$ value is estimated via the following function:[44,45]

$$S_R = \left|\frac{1}{\tau}\frac{d\tau}{dT}\right| \times 100\% \quad (8)$$

Accordingly, the calculated $S_r$ value as a function of temperature is illustrated in Figure 3(i). With arising the temperature, it is found that the $S_R$ value increases monotonously, achieving its maximum value of 0.86% K$^{-1}$ at 623 K. Clearly, the obtained $S_r$ value is comparable with previously proposed lifetime-based optical thermometers, such as Ba$_2$GdV$_3$O$_{11}$:Er$^{3+}$/Yb$^{3+}$ ($S_R$ = 1.1% K$^{-1}$), Cs$_2$PtCl$_6$ ($S_R$ = 1.21% K$^{-1}$), CaYGaO$_4$:Cr$^{3+}$ ($S_R$ = 0.78% K$^{-1}$), Y$_2$Mo$_4$O$_{15}$:Er$^{3+}$/Yb$^{3+}$



($S_R$ = 0.39 K$^{-1}$), *etc.*,[44,45-48] demonstrating that the designed phosphors can be used as lifetime-based optical thermometers. These achievements confirm that the Mn$^{2+}$-doped Mg$_2$Y$_8$(SiO$_4$)$_6$O$_2$ phosphors are promising candidates for multi-parameter optical thermometry, *i.e.*, using the emission band centroid and lifetime as the thermometric parameters.

To study the evolution of the phase structure of the final products at high-pressure, the *in-situ* high-pressure XRD patterns of the Mg$_2$Y$_8$(SiO$_4$)$_6$O$_2$:0.14Mn$^{2+}$ phosphors were examined and demonstrated in Figure 4(a). Here, since the high energy of Mo *Kα* irradiation (λ = 0.7107 Å) was employed to execute the *in-situ* XRD measurement, the positions of the diffraction peaks are inconsistent with those of the results presented in Figure 1(a). Obviously, at high-pressure, neither do the diffraction peaks vanish nor do new diffraction peaks from secondary phase generate, stating that the pressure-induced phase transition does not happen in the studied samples within the applied range of 0.27-10.90 GPa. Nevertheless, it can be seen that the diffraction peaks are simultaneously broadened and shift to larger angles (see Figure 4(a)). The pressure-induced strains as well as the generation of some crystal defects in the compressed structure can be responsible for the broadened diffraction peaks at high-pressure. Moreover, the shift of the diffraction peaks is assigned to the pressure-triggered lattice contraction, which contributes to the adjusting the crystal field surrounding the emitting center, *i.e.*, Mn$^{2+}$, resulting in the evident spectral shift at high-pressure and it will be analyzed in the following section. Note that, when the designed compounds undergo the decompression process, the diffraction peaks can go back to their starting states (*i.e.*, shifting to the smaller angles) (Figure 4(a)), suggesting reversibility of the observed changes confirming high structure stability of the synthesized phosphors. For the aim of describing the pressure-dependent crystal structure in depth, the Rietveld XRD refinements of the diffractograms has been performed and the corresponding results are illustrated in Figure 4(b), 4(c) and S4. As disclosed, these calculated diffraction peaks are identical with the experimental data, manifesting that the synthesized phosphors possess pure hexagonal phase, which is hardly impacted by high-pressure. In



addition, as pressure rises, a gradual decrement is seen in the lattice parameters, *i.e.*, $a = b$, $c$ and volume of the unit cell of the studied samples (Figure 4(d) and 4(e)). Particularly, the volume of the unit cell of $Mg_2Y_8(SiO_4)_6O_2:0.14Mn^{2+}$ phosphors decreases from 485.522 to 446.939 Å$^3$ as the pressure raises from 0.27 to 10.90 GPa (see Figure 4(e)). These results reveal that the $Mn^{2+}$-doped $Mg_2Y_8(SiO_4)_6O_2$ phosphors present splendid phase structure stability throughout the interested pressure range, which ensures their usable in high-pressure conditions.

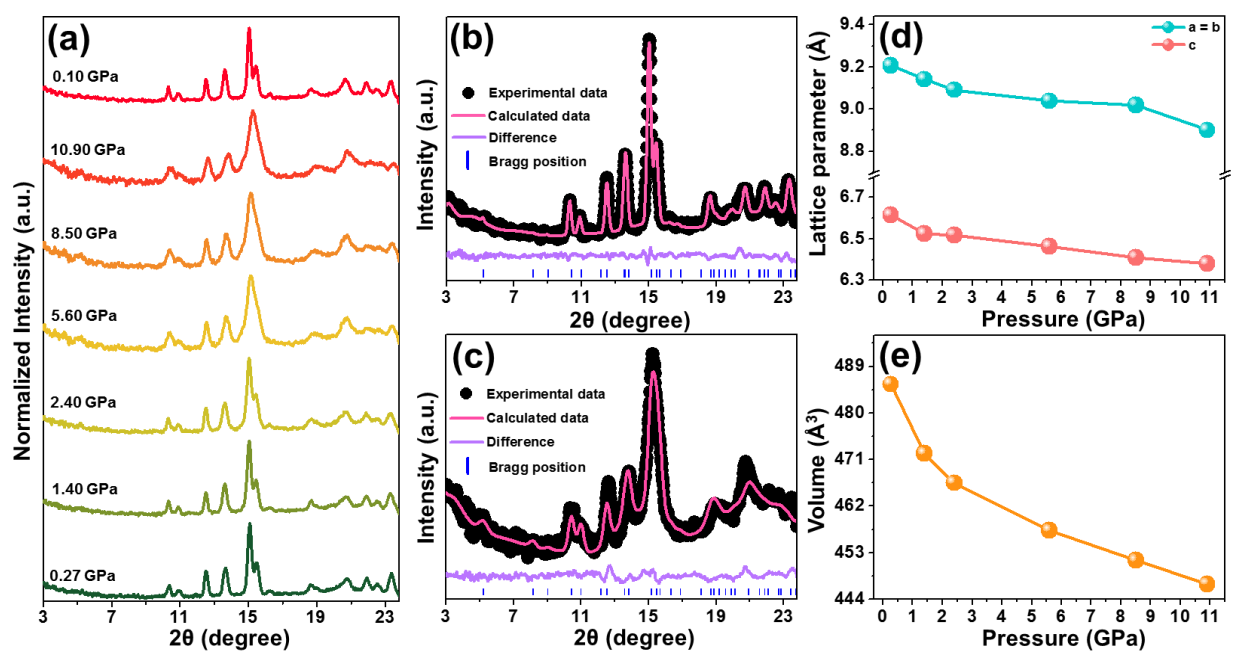

**Figure 4** (a) *In-situ* high-pressure XRD profiles of the $Mg_2Y_8(SiO_4)_6O_2:0.14Mn^{2+}$ phosphors. Rietveld XRD refinements of the $Mg_2Y_8(SiO_4)_6O_2:0.14Mn^{2+}$ phosphors recorded at (b) 1.40 GPa and (c) 10.90 GPa. (d) Lattice parameter (*i.e.*, $a = b$, $c$) and (e) cell volume of the studied samples as a function of pressure.

As is well documented, the luminescence features of $Mn^{2+}$ originate from *d-d* electronic transition, which is greatly decided by its surrounding crystal field, endowing its *in-situ* regulation at high-pressure. Here, to analyze the spectral evolution of the studied samples at high-pressure, the *in-situ* high-pressure measurement was carried out and the corresponding experimental setup is described in Figure 5(a). Upon 365 nm excitation, the pressure-dependent emission spectra of the $Mg_2Y_8(SiO_4)_6O_2:0.14Mn^{2+}$ phosphors in the range of 0.13-10.89 GPa were tested (Figure 5(b)). Although the high-pressure is applied, the resulting phosphors can



still exhibit the luminescence characteristics of $Mn^{2+}$. Nevertheless, an evident spectral red-shift is gained in the designed compounds, of which the emission band centroid alters from 613.8 to 656.7 nm (*i.e.*, a displacement of 42.9 nm) with increasing the pressure from 0.13 to 10.89 GPa. Because of the obvious spectral shift, high-pressure induced color-tunable luminescence is realized in the resultant phosphors, namely, the emitting-color changes from orange to red and the corresponding color CIE 1931 chromatic coordinates shift from (0.563,0.432) to (0.638,0.342) within the applied pressure range (see Figure 5(c) and Table S2). Furthermore, it is demonstrated in Figure S5 that the spectral features of the synthesized phosphors can return to their initial states (*i.e.*, spectral blue-shift) when they experiences the decompression process. This result implies that the $Mn^{2+}$-doped $Mg_2Y_8(SiO_4)_6O_2$ phosphors present remarkable structure stability and reversibility, which makes their promising applications in high-pressure conditions. As is known, with rising the pressure, the lattice contraction takes place, which will shorten the bond length, and thus, the energy of the emitting level of the $Mn^{2+}$ will be decreased at high-pressure, resulting in the pressure-caused spectral red-shift. Additionally, the ionic distance between the $Mn^{2+}$ and $O^{2-}$ shortens during the compression process, which is beneficial for reinforcing the covalent characteristics owing to the improved nephelauxetic effect. As a result, the energy separation between the ground state and excited level will be reduced, contributing to the spectral red-shift at high-pressure. Due to the combined effect of these aforementioned factors, the pressure-associated red-shift of the emission band centroid is observed in the designed phosphors.

Dependence of the emission band centroid, which was evaluated from the pressure-dependent emission spectra, on the pressure is depicted in Figure 5(d). Clearly, a polynomial function can be employed to effectively fit the pressure-dependent emission band centroid and it is expressed as: $\lambda = 613.70 + 3.01p + 0.08p^2$. Thus, the pressure sensitivity (*i.e.*, $d\lambda/dp$) of the $Mg_2Y_8(SiO_4)_6O_2:0.14Mn^{2+}$ phosphors is obtained and presented in Figure 5(e). As disclosed, the $d\lambda/dp$ value increases monotonously as pressure increases, reaching up to its maximum



value of 4.75 nm GPa$^{-1}$. Note that, this obtained d$\lambda$/d$p$ value is 18.6 and 13.0 times higher than the commonly used optical manometers of SrB$_4$O$_7$:Sm$^{2+}$ (d$\lambda$/d$p$ = 0.255 nm GPa$^{-1}$) and ruby (d$\lambda$/d$p$ = 0.365 nm GPa$^{-1}$), respectively. Moreover, to further clarify the manometric characteristics of the studied samples, we summarized pressure sensing capacities of the previously developed optical manometers that operated in the visible light range, as listed in Table 1. Compared with the previously developed optical manometers (Table 1), the designed phosphors do not only exhibit superior pressure sensitivity but also possess relative broad operating range, suggesting that the Mn$^{2+}$-doped Mg$_2$Y$_8$(SiO$_4$)$_6$O$_2$ phosphors are promising candidates for highly-sensitive optical manometry via adopting the emission band centroid as the pressure indicator.

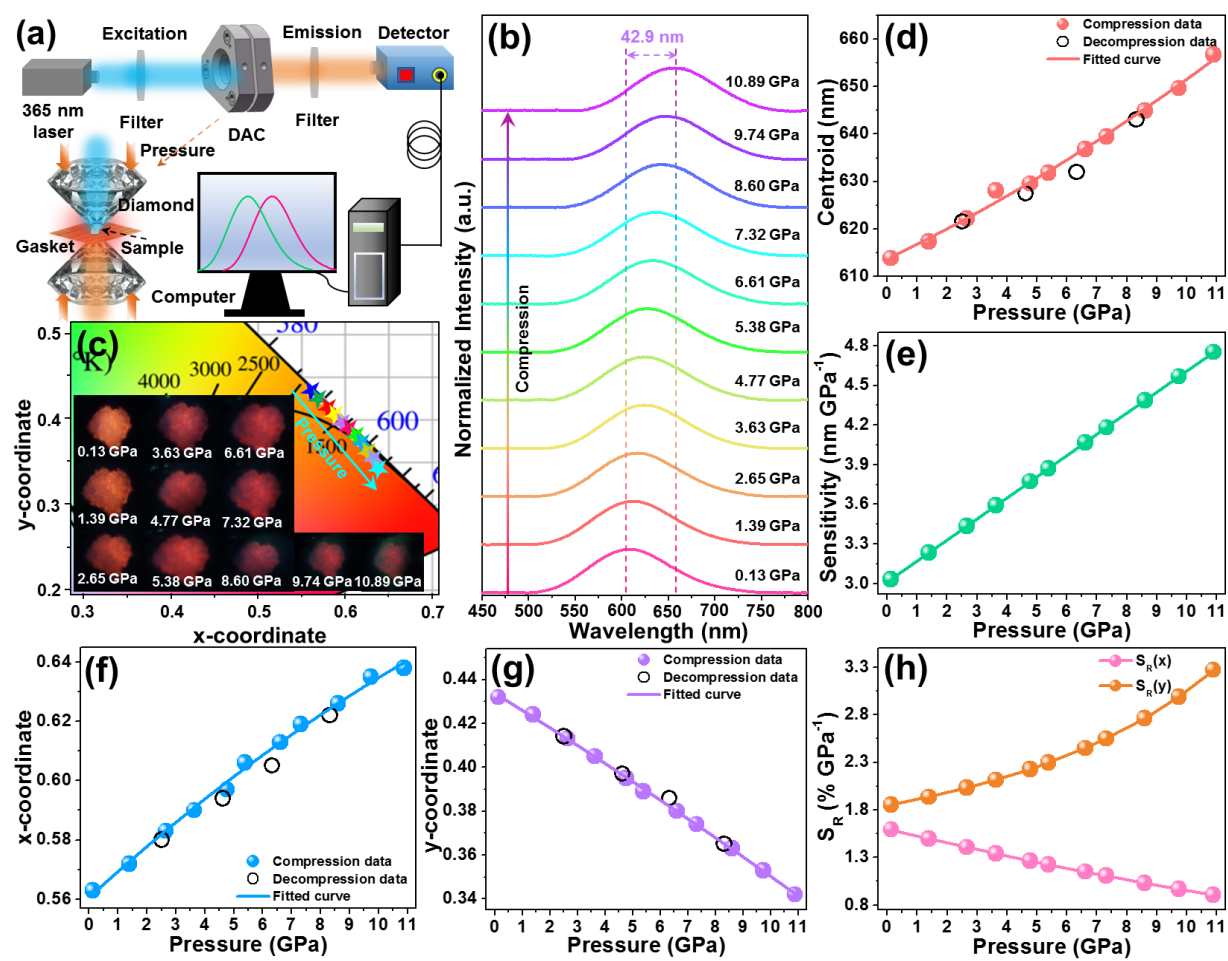

**Figure 5** (a) Experimental setup for high-pressure measurement. Pressure-related (b) emission spectra ($\lambda_{exc}$ = 365 nm), (c) CIE chromaticity, (d) Emission band centroid, (e) Sensitivity (*i.e.*, d$\lambda$/d$p$), (f) *x*-coordinate,



(g) $y$-coordinate and (h) $S_r$ value of the Mg$_2$Y$_8$(SiO$_4$)$_6$O$_2$:0.14Mn$^{2+}$ phosphors. Inset of (c) shows the optical images of luminescence of Mg$_2$Y$_8$(SiO$_4$)$_6$O$_2$:0.14Mn$^{2+}$ at high-pressure.

**Table 1** Comparison of the manometric properties of the designed phosphors with previously reported bandshift-based optical manometers emitting in visible spectral range.

| Compounds | Operating range | Centroid | Sensitivity | Reference |
|---|---|---|---|---|
| Ca$_8$Zn(SiO$_4$)$_4$Cl$_2$:Eu$^{2+}$ | 0.46-20.35 GPa | 500.2 nm | 4.18 nm GPa$^{-1}$ | 6 |
| Li$_4$SrCa(SiO$_4$)$_2$:Eu$^{2+}$ | 0-15.69 GPa | 584 nm | 5.19 nm GPa$^{-1}$ | 11 |
| Sr$_2$[Mg$_{0.9}$Li$_{0.1}$Al$_{4.9}$Si$_{0.1}$N$_7$]:Eu$^{2+}$ | 0-10.20 GPa | 485 nm | 3.51 nm GPa$^{-1}$ | 12 |
| Ba$_3$Lu(BO$_3$)$_3$:Ce$^{3+}$ | 0-20.8 GPa | 485 nm | 3.51 nm GPa$^{-1}$ | 13 |
| NaMgBO$_3$:Ce$^{3+}$ | 0.31-17.8 GPa | 469.3 nm | 2.94 nm GPa$^{-1}$ | 14 |
| Zn$_3$Ga$_2$GeO$_8$:Mn$^{4+}$ | 1.4-19.9 GPa | ~696.5 nm | 0.844 nm GPa$^{-1}$ | 15 |
| La$_3$Mg$_2$SbO$_9$:Mn$^{4+}$ | 0-9.48 GPa | ~690.5 nm | 1.2 nm GPa$^{-1}$ | 16 |
| Ca$_2$Gd$_8$(SiO$_4$)$_6$O$_2$:Mn$^{2+}$ | 0.38-13.66 GPa | 592.2 nm | 7.25 nm GPa$^{-1}$ | 27 |
| Mg$_2$Y$_8$(SiO$_4$)$_6$O$_2$:Mn$^{2+}$ | 0.13-10.89 GPa | 613.8 nm | 4.75 nm GPa$^{-1}$ | This work |

From Figure 5(c) and Table S2, it is clear that the color coordinate of the resulting phosphors is highly dependent on pressure. Moreover, it is well known that the change degree of the color coordinate can be quantitatively characterized by the chromaticity shift, *i.e.*, Δs, as defined below:[6,42]

$$\Delta s = \sqrt{(u'_m - u'_0)^2 + (v'_m - v'_0)^2 + (w'_m - w'_0)^2} \quad (9)$$

$$w' = 1 - v' - w' \quad (10)$$

$$v' = 9y/(3 - 2x + 12y) \quad (11)$$

$$u' = 9y/(3 - 2x + 12y) \quad (12)$$

where the color coordinates of the studied samples at initial and measured pressure are labelled by 0 and *m*, respectively. Accordingly, it is found that the Δs value of the prepared phosphors



increases gradually as pressure elevates, that is, it improves from 0.05 to 0.21 (see Figure S6). Since the shift of the color coordinate is significant, remote optical manometry is also expected to be realized by analyzing the pressure-dependent color coordinate. The *x*-coordinate and *y*-coordinate of the $Mg_2Y_8(SiO_4)_6O_2$:0.14$Mn^{2+}$ phosphors are depicted in Figure 5(g) and 5(h), respectively. Herein, a 2rd-order polynomial equation is able to be employed to fit the pressure-related color coordinate, *i.e.*, $x = -1.466 \times 10^{-4}p^2 + 0.09p + 0.56$ and $y = -5.481 \times 10^{-5}p^2 - 0.008p + 0.43$. For the sake of investigating the optical pressure sensing properties of the studied samples, the relative pressure sensitivity (*i.e.*, $S_R$) is required to be clarified, as described below:[6,47]

$$S_R = \left|\frac{1}{C}\frac{dC}{dT}\right| \times 100\% \tag{13}$$

where *C* stands for the chromatic coordinate and the relative pressure sensitivities, which are estimated from *x*-coordinate and *y*-coordinate, are denoted by $S_R(x)$ and $S_R(y)$, respectively. Figure 5(h) shows the $S_R(x)$ and $S_R(y)$ of the $Mg_2Y_8(SiO_4)_6O_2$:0.14$Mn^{2+}$ phosphors as a function of pressure. As demonstrated, as pressure increases, the $S_R(x)$ value decreases gradually and its maximum value is 1.59% $GPa^{-1}$, whereas the $S_R(y)$ shows an opposite changing tendency, *i.e.*, it increases monotonously, of which its maximum value is 3.27% $GPa^{-1}$, demonstrating that the resulting phosphors can be used as colorimetric pressure sensing. These results manifest that the $Mn^{2+}$-doped $Mg_2Y_8(SiO_4)O_2$ phosphors with good pressure sensitivities are promising pressure-sensitive luminescent materials for multi-parameter optical manometry, *i.e.*, using the emission band centroid and color coordinate as the pressure indicators.

The pressure-induced spectral shift of the $Mn^{2+}$ emission band in $Mg_2Y_8(SiO_4)_6O_2$:0.14$Mn^{2+}$, as described above, can be effectively utilized for pressure sensing. However, this approach is not practical for two-dimensional pressure mapping, as it requires point-by-point acquisition of emission spectra, which is extremely time-consuming. As an alternative, a ratiometric method, recently introduced, can be implemented, based on the



intensity ratio of the same $Mn^{2+}$ emission band measured across two distinct spectral regions (see Figure 6(a)). This enables rapid pressure imaging. By analyzing the pressure-dependent emission intensity of $Mn^{2+}$ in four selected spectral intervals (Figure 6b), it is evident that the red-shift of the emission band leads to a more than 14-fold and 10-fold decrease in intensity for regions $I_1$ and $I_2$, respectively, over the investigated pressure range. Simultaneously, a 4-fold and 5-fold increase in emission intensity is observed for $I_3$ and $I_4$. These opposite monotonic pressure dependencies allow the definition of ratiometric pressure parameters in the form of luminescence intensity ratios (*LIRs*). The *LIR* vs. pressure plots shown in Figure 6(c) demonstrate clear monotonic trends across the entire pressure range. The manometric relative sensitivities, calculated using the following equation:[19,20]

$$S_R = \frac{1}{LIR}\frac{\Delta LIR}{\Delta p} \times 100\% \qquad (14)$$

The calculated $S_R$ values are shown in Figure 6(d), of which the highest values are obtained when the $LIR_2$ is used as the manometric parameter, *i.e.*, yielding $S_R = 56\%$ GPa$^{-1}$ at 2.8 GPa and 72% GPa$^{-1}$ at 7 GPa. Notably, above 8 GPa, the relative sensitivities obtained for $LIR_3$ and $LIR_4$ significantly exceed those for $LIR_2$. This behavior suggests that, using the same luminescent manometer and merely adjusting the spectral regions analyzed, the pressure operating range for highly sensitive pressure readout can be extended. From the perspective of practical pressure sensing, it is crucial that the manometric parameter remains insensitive to temperature variations; otherwise, the reliability of the pressure readout is compromised. Given that the $Mn^{2+}$ emission band does not exhibit significant spectral shifts with temperature, an expected and favorable outcome, the proposed ratiometric method should allow pressure sensing unaffected by thermal fluctuations.



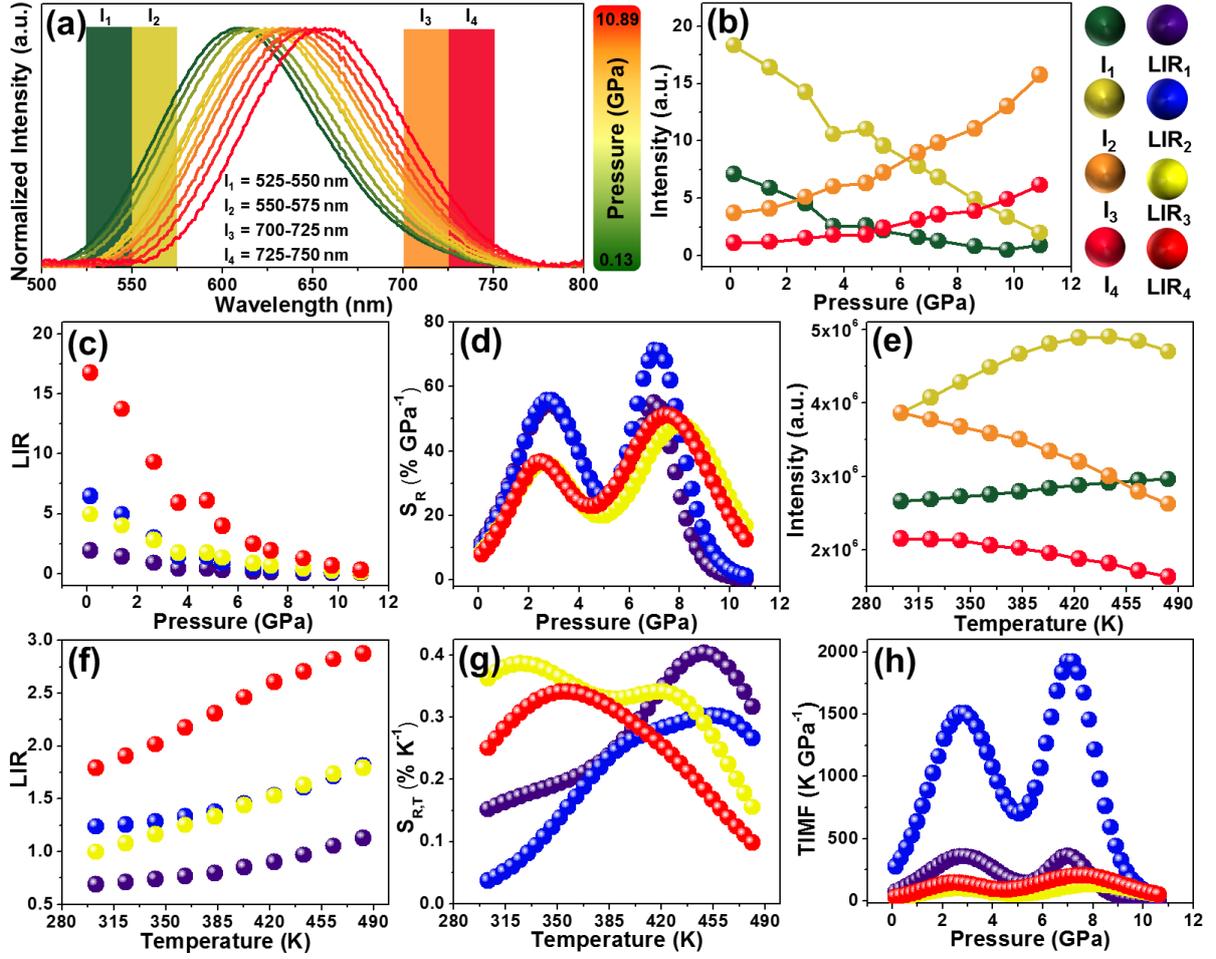

**Figure 6** (a) Normalized emission spectra, (b) Integrated emission intensities (*i.e.*, $I_1$, $I_2$, $I_3$ and $I_4$), (c) LIR values and (d) $S_r$ values of the $Mg_2Y_8(SiO_4)_6O_2$:0.14$Mn^{2+}$ phosphors as a function of pressure. Temperature-dependent (e) Integrated emission intensities (*i.e.*, $I_1$, $I_2$, $I_3$ and $I_4$), (f) LIR values and (g) $S_{R,T}$ values of the $Mg_2Y_8(SiO_4)_6O_2$:0.14$Mn^{2+}$ phosphors. (h) TIMF values of the $Mg_2Y_8(SiO_4)_6O_2$:0.14$Mn^{2+}$ phosphors.

To verify this hypothesis, the temperature dependence of emission intensities $I_1$-$I_4$ was analyzed. As shown in Figure 6(e), only minor variations in the relative emission intensities from $Mn^{2+}$ ions occupying different crystallographic sites were observed, resulting in small changes in the corresponding *LIRs* over the examined temperature range. The largest variation was found for *LIR$_4$*, which doubled in value between 300 K and 500 K (Figure 6(f)). However, in the same temperature range, *LIR$_2$* remained stable up to approximately 375 K, with a subsequent increase of about 28% at higher temperature. This translates into low thermal relative sensitivities for all *LIR* parameters (Figure 6(g)), none of which exceeded 0.4% $K^{-1}$



across the full temperature range. Specifically, $LIR_2$ exhibited a thermal relative sensitivity of only ~0.038% K$^{-1}$ at room temperature, which is remarkably low. To quantify the thermal robustness of the pressure readout, the temperature-independent manometric factor (TIMF) was calculated:[19,20]

$$\text{TIMF} = \left|\frac{S_{R,P,Max}}{S_{R,T,(RT)}}\right| \quad (15)$$

where the maximum pressure relative sensitivity is presented by $S_{R,P,Max}$ and the temperature sensitivity achieved at room temperature is labeled by $S_{R,T,(RT)}$. Due to instrumental limitations, this calculation was based on the thermal relative sensitivity at room temperature (300 K). As shown in Figure 6(h), the combination of low thermal sensitivity and high pressure-induced variability of $LIR_2$ resulted in exceptionally high TIMF values, with TIMF$_{max}$ reaching 1940 K GPa$^{-1}$ at 7 GPa. A comparison of manometric parameters with other luminescent manometers using a similar approach (Table 2) indicates that only the Cr$^{3+}$-doped MgO phosphors has demonstrated a higher TIMF value of 3800 K GPa$^{-1}$. However, it is particularly noteworthy that for all previously reported ratiometric luminescent manometers, such high TIMF values have only been achieved in relatively low-pressure ranges (< 3 GPa). Therefore, the fact that the designed materials exhibit such high TIMF values above 3 GPa clearly positions it as a highly promising candidate for advanced pressure sensing applications.

**Table 2** Comparison of manometric performance of the different ratiometric luminescence manometers

| Compounds | $S_{R,p\ max}$ | $p@S_{R,pmax}$ | TIMF | Reference |
|---|---|---|---|---|
| Ca$_{0.8}$Sr$_{0.2}$MgSi$_2$O$_6$:Cr$^{3+}$ | 50.7% Gpa$^{-1}$ | ambient | 1558 K GPa$^{-1}$ | 5 |
| Li$_2$Mg$_3$TiO$_6$:Cr$^{3+}$ | 4.7% Gpa$^{-1}$ | ambient | 62.6 K GPa$^{-1}$ | 17 |
| MgO:Cr$^{3+}$ | 9.83% GPa$^{-1}$ | 1.15 GPa | 3800 K GPa$^{-1}$ | 18 |
| Li$_3$Sc$_2$(PO$_4$)$_3$:Cr$^{3+}$ | 56.86% GPa$^{-1}$ | 0.02 GPa | 1386.8 K GPa$^{-1}$ | 19 |
| LiScGeO$_4$:Cr$^{3+}$ | 120% GPa$^{-1}$ | 2.55 GPa | 241 K GPa$^{-1}$ | 49 |
| CaAl$_{12}$O$_{19}$:Cr$^{3+}$ | 70% GPa$^{-1}$ | ambient | 150 K GPa$^{-1}$ | 50 |
| MgGeO$_3$:Cr$^{3+}$ | 62% GPa$^{-1}$ | 4.9 GPa | 600 K GPa$^{-1}$ | 51 |
| Mg$_2$Y$_8$(SiO$_4$)$_6$O$_2$:Mn$^{2+}$ | 72% GPa$^{-1}$ | 7 GPa | 1940 K GPa$^{-1}$ | This work |

## 3. Conclusion

In summary, series of the Mn$^{2+}$-doped Mg$_2$Y$_8$(SiO$_4$)$_6$O$_2$ phosphors were prepared through a high-temperature solid-state reaction method. Under 408 nm excitation, intense orange emission from Mn$^{2+}$ is observed in the resultant phosphors and the concentration quenching takes place when $x > 0.07$, contributed by the electric quadrupole-quadrupole interaction. The synthesized phosphors possess splendid thermal stability with an activation energy of 0.255 eV, as verified by the temperature-related emission spectra. Through analyzing the temperature-dependent emission band centroid and lifetime, the temperature sensitivities of the Mg$_2$Y$_8$(SiO$_4$)$_6$O$_2$:0.14Mn$^{2+}$ phosphors are found to be $d\lambda/dT = 0.053$ nm K$^{-1}$ and $S_R = 0.86\%$ K$^{-1}$, respectively. The *in-situ* high-pressure XRD patterns confirm that the prepared phosphors have high structural stability and reversibility within the pressure range of 0.27-10.90 GPa, ensuring their feasibilities in high-pressure environment. Owing to the pressure-induced contraction structure, evident spectral red-shift is observed in the studied samples, resulting in the color-tunable luminescence at high-pressure. Through analyzing the pressure-dependent emission band centroid, one knows that the pressure sensitivity of the designed phosphors is $d\lambda/dp = 4.75$ nm GPa$^{-1}$ and its operating ranging is 0.13-10.89 GPa. Furthermore, utilizing the color coordinate as the pressure indicator, the manometric properties of the synthesized phosphor are also investigate, in which the $S_R(x)$ and $S_R(y)$ values are 1.59% and 3.27% GPa$^{-1}$, respectively. Moreover, low thermal variation of the spectral position of Mn$^{2+}$ ions emission band and its high pressure variability enables the development of highly sensitive ratiometric luminescence manometers with the $S_{R,pmax} = 72$ % GPa$^{-1}$ at 7 GPa and the highest up to date TIMF reported above 3 GPa reaching 1940 K GPa$^{-1}$ at 7 GPa. These results suggest that the Mn$^{2+}$-doped Mg$_2$Y$_8$(SiO$_4$)$_6$O$_2$ phosphors with proper luminescence and sensing characteristics are potential bifunctional platforms for multi-parameter optical manometry and thermometry.

## 4. Experimental Section

*Preparation of Mn$^{2+}$-doped ZnAl$_2$O$_4$ phosphors:* The Mg$_{2-2x}$Y$_8$(SiO$_4$)$_6$O$_2$:2$x$Mn$^{2+}$ (*i.e.*, Mg$_2$Y$_8$(SiO$_4$)$_6$O$_2$:2$x$Mn$^{2+}$, $0.01 \leq x \leq 0.11$) phosphors were synthesized via the high-temperature solid-state reaction method. According to the designed compounds, the proper amounts of the powders, *i.e.*, MgCO$_3$, Y$_2$O$_3$, SiO$_2$ and MnCO$_3$, were weighted and mixed thoroughly via an agate mortar. The mixed raw materials were then put in crucibles and kept in a furnace, of which the sintering temperature and time were 1450 °C and 6 h, respectively. When the temperature naturally cooled down, the final products were collected and reground for further characterization.

*Measurement and characterization:* The impact of the Mn$^{2+}$ doping on the phase structure of the final products was analyzed by an X-ray diffractometer (Bruker D8 Advance, Cu K$\alpha$ irradiation). The elemental compositions as well as their valences were examined by an X-ray photoelectron spectrometer. The diffuse reflectance spectra of designed compounds were detected through an UV-Vis spectrophotometer (Cary 500). The morphological performances of the resulting phosphors were checked via a transmission electron microscope (TEM; JEM-2100F, JEOL) and a scanning electron microscope (SEM; HITACHI SU3500), of which an energy-dispersive X-ray spectroscopy (EDS) spectrometer was attached. Through an electron paramagnetic resonance (EPR) spectrometer (Bruker EMXplus-6/1), the EPR analysis was carried out. The luminescence features, *i.e.*, excitation and emission spectra, of the developed phosphors were measured by a fluorescence spectrometer (Edinburgh FS5). Furthermore, the room-temperature decay curves of the resulting phosphors were tested by means of a fluorescence spectrometer (ZoLix Omni-λ3005i). The luminescence decay curves were measured using the FLS1000 Fluorescence Spectrometer from Edinburgh Instruments equipped with a 150 W μFlash lamp and R928 photomultiplier tube from Hamamatsu. The temperature of the sample was controlled by using a THMS600 heating - cooling stage from Linkam (0.1 K temperature stability and 0.1 K point resolution).

With an aid of a DAC, the *in-situ* high-pressure measurements were conducted. As for the gasket, the T301 steel was employed, which was pre-indented to 40 μm and drilled a hole (*i.e.*, 150 μm). Then, the designed phosphors and ruby were put into the hole. The pressure in DAC was calibrated by analyzing the luminescence of ruby. The pressure-related emission spectra were measured via a fluorescence spectrometer (MicroHP-SP), in which a 365 nm semiconductor laser was adopted as the excitation lighting source. *In-situ* high pressure angle-dispersive XRD measurement was performed with Mo radiation source (Mo-*Kα*, $\lambda$ = 0.7107 Å, Rigaku Nanopix WE). The *in-situ* high-pressure XRD measurements were carried out at room temperature. The theoretical calculation details of the electronic structure are presented in the Supporting Information.

**Acknowledgements**

This work was supported by Zhejiang Provincial Natural Science Foundation of China (LMS25F050008).

**Competing interests**

The authors declare no competing financial interest.

**References**

[1] Y. Zhou, G. Ledoux, L. Bois, G. Pilet, M. Colombo, E. Jeanneau, L. Lafarge, C. Journet, S. Descartes, D. Philippon, *Adv. Optical Mater*. **2024**, *12*, 2301800.

[2] L. Marciniak, P. Wozny, M. Szymczak, M. Runowski, *Coord. Chem. Rev*. **2024**, 507, 215770.

[3] D. Li, M. Jia, T. Jia, G. Chen, *Adv. Mater*. **2024**, *36*, 2309452.

[4] Y. Tong, W. Liu, S. Ding, *J. Rare Earths*. **2024**, *42*, 1507-1513.

[5] M. Szymczak, K. Su, L. Mei, M. Runowski, P. Woźny, Q. Guo, L. Liao, L. Marciniak, *ACS Appl. Mater. Interfaces* **2024**, *16*, 60491-60500.

[6] P. Du, J. Xue, S. Ma, P. Woźny, V. Lavín, L. Luo, M. Runowski, *Mater. Horiz*. **2025**, *12*, 4822-4832.

[7] W. Zhang, J. Wang, L. Wang, J. Wan, G. Bai, S. Xu, L. Chen, *Adv. Optical Mater*. **2025**, *13*, 2403290.

[8] P. Wu, T. Zhang, Y. Ding, H. Yang, F. Zhao, Q. Mao, M. Liu, J. Zhong, *Inorg. Chem*. **2025**, *64*, 6713-6721.

[9] L. Zhou, R. Chen, X. Li, M. Chang, Y. Huo, Y. Yang, C. Li, W. Yang, H. Lin, L. Liu, S. Li, F. Zeng, *ACS Appl. Mater. Interfaces* **2025**, *17*, 25451-25466.

[10] L. Pu, P. Li, J. Zhao, Y. Wang, D. Guo, L. Li, Z. Wang, H. Suo, *Laser Photonics Rev*. **2023**, *17*, 220884.

[11] K. Su, L. Mei, Q. Guo, P. Shuai, Y. Wang, Y. Liu, Y. Jin, Z. Peng, B. Zou, L. Liao, *Adv. Funct. Mater*. **2023**, *33*, 2305359.

[12] Z. Li, H. Wang, Z. Su, R. Kang, T. Seto, Y. Wang, *Angew. Chem. Int. Edit*. **2025**, *137*, e202419910.

[13] Y. Zhao, Z. Zheng, Z. Li, Z. Shi, Y. Song, B. Zou, H. Zou, *Inorg. Chem*. **2024**, *63*, 4288-4298.

[14] Y. Ma, W. Meng, M. Runowski, V. Lavín, U. R. Rodríguez-Mendoza, J. H. Jeong, J. Xue, *J. Alloys Compd*. **2024**, *1044*, 184350.

[15] C. Su, D. Xu, K. Li, X. Xie, Y. Ren, Y. Liu, Y. Jin, M. Du, P. Shen, *J. Phys. Chem. C* **2024**, *128*, 16791-16796.

[16] Z. Chen, S. Du, F. Li, S. Zhang, S. Zhao, Z. Tian, J. Zhang, X. Yuan, G. Liu, K. Chen, *J. Mater. Sci. Technol*. **2024**, *194*, 98-109.

[17] M. Szymczak, P. Woźny, M. Runowski, M. Pieprz, V. Lavín, L. Marciniak, *Chem. Eng. J.* **2022**, *453*, 139632.

[18] M. Szymczak, M. Runowski, V. Lavín, L. Marciniak, *Laser Photonics Rev.* **2023**, *17*, 2200801.

[19] M. Szymczak, J. Jaskielewicz, M. Runowski, J. Xue, S, Mahlik, L. Marciniak, *Adv. Funct. Mater.* **2024**, *34*, 2314068.

[20] M.Szymczak, P. Du, M. Runowski, P. Woźny, J. Xue, T. Zheng, L. Marciniak, *Adv. Opt. Mater* **2023**, *12*, 2302147.

[21] H. Li, Y. Zhang, L. Li, M. Shen, G. Niu, Y. Xu, M. Xia, *Chem. Eng. J.* **2025**, *522*, 167583.

[22] L. Yan, G. Zhu, S. Ma, S. Li, Z. Li, X. Luo, B. Dong, *Laser Photonics Rev.* **2024**, *18*, 2301200.

[23] Y. Tang, Y. Cai, K. Dou, J. Chang, W. Li, S. Wang, M. Sun, B. Huang, X. Liu, J. Qiu, L. Zhou, M. Wu, J. C. Zhang, *Nat. Commun.* **2024**, *15*, 3209.

[24] G. Lu, Y. Wang, K. Ma, X. Chen, W. Geng, T. Liu, S. Xu, J. Zhang, B. Chen, *Ceram. Int.* **2024**, *50*, 16190-16200.

[25] H. Liu, Y. Shao, C. Dou, J. Zhao, Z. Song, Q. Liu, *Adv. Optical Mater.* **2025**, *13*, 2403472.

[26] C. Zhan, H. Zhu, S. Liang, W. Nie, Z. Wang, M. Hong, *Adv. Optical Mater.* **2024**, *12*, 2400574.

[27] P. Du, J. Xue, A. M. González, L. Luo, P. Woźny, U. R. Rodríguez-Mendoza, V. Lavín, M. Runowski, *Chem. Eng. J.* **2025**, *505*, 159652.

[28] Z. Zheng, Z. Li, H. Zou, Q. Tao, Y. Zhao, Q. Wang, Z. Shi, Y. Song, L. Li, *Inorg. Chem.* **2024**, *63*, 3882-3892.

[29] X. Xiao, Z. Li, H. Zou, Q. Sun, Y. Song, Q. Tao, L. Li, B. Zou, *Inorg. Chem.* **2024**, *63*, 5175-5184.

[30] W. Y. Li, Y. F. Lei, K. T. Wu, M. Xu, W. B. Dai, *Ceram. Int.* **2024**, *50*, 48021-48030.

[31] N. Hussain, S. Rubab, V. Kumar, *Ceram. Int.* **2023**, *49*, 15341-15348.

[32] J. Lin, Q. Su, *J. Mater. Chem*. **1995**, *5*, 1151-1154.

[33] Q. Su, J. Lin, B. Li, *J. Alloys Compd*. **1995**, *225*, 120-123.

[34] G. Li, D. Geng, M. Shang, Y. Zhang, C. Peng, Z. Cheng, J. Lin, *J. Phys. Chem. C* **2011**, *115*, 21882-21892.

[35] C. Yang, J. Long, B. Li, R. Ma, B. Cao, C. Deng, W. Huang, *J. Alloys Compd*. **2025**, *1010*, 177035.

[36] X. Huang, T. Wang, W. Zhang, Z. Zhao, X. Wang, X. Gao, *ACS Appl. Electron. Mater*. **2021**, *3*, 4472-4483.

[37] S. Banerjee, S. Saikia, M. S. Molokeev, A. Nag, *Chem. Mater*. **2024**, *36*, 4750-4757.

[38] D. L. Dexter, *J. Chem. Phys*. **1953**, *21*, 836–850.

[39] M. Wei, Y. Liang, J. Zeng, T. Zhao, Y. Hao, X. Zhang, W. Li, H. Zhang, B. Lei, *ACS Appl. Mater. Interfaces* **2025**, *17*, 32657-32666.

[40] Z. Yang, G. Lu, J. Ma, T. Yang, G. Xiang, L. Li, X. Zhou, Z. Xia, *Laser Photonics Rev*. **2025**, *19*, 2500300.

[41] Q. Zeng, M. Runowski, J. Xue, L. Luo, L. Marciniak, V. Lavín, P. Du, *Adv. Sci*. **2024**, *11*, 2308221.

[42] Y. Huang, Y. Pan, S. Guo, C. Peng, H. Lian, J. Lin, *Inorg. Chem*. **2022**, *61*, 8356-8365.

[43] M. Yu, D. Zhao, R. Zhang, Q. Yao, L. Jia, Q. Zong, *Ceram. Int*. **2024**, *50*, 11766-11775.

[44] I. Kachou, K. Saidi, U. Ekim, M. Dammak, M. Ç. Ersundu, A. E. Ersundu, *Dalton Trans*. **2024**, *53*, 2357-2372.

[45] M. Szymczak, W. M. Piotrowski, P. Woźny, M. Runowski, L. Marciniak, *J. Mater. Chem. C* **2024**, *12*, 6793-6804.

[46] Z. Pei, M. Runowski, X. Ge, P. Woźny, L. Luo, P. Du, *ACS Appl. Mater. Interfaces* **2025**, *17*, 49683-49691.

[47] Z. Pei, M. Runowski, P. Woźny, P. Du, N. Chen, *Inorg. Chem*. **2025**, *64*, 20439-20448.

[48] K. Saidi, C. Hernández-Álvarez, M. Runowski, M. Dammak, I. R. Martín, *Dalton Trans*. **2023**, *52*, 14904-14916.

[49] M. Szymczak, M. Runowski, M.G. Brik, L. Marciniak, *Chem. Eng. J.* **2023**, *466*, 143130.

[50] M. Szymczak, A. Antuzevics, P. Rodionovs, M. Runowski, U. R. Rodríguez-Mendoza, D. Szymanski, V. Kinzhybalo, L. Marciniak, *ACS Appl. Mater. Interfaces* **2024**, *16*, 64976-64987.

[51] M. Szymczak, W. M. Piotrowski, U. R. Rodríguez-Mendoza, P. Wozny, M. Runowski, L. Marciniak, *J. Mater. Chem. C* **2025**, *13*, 4224-4235.